\documentclass[preprintnumbers,prd,showpacs,floatfix,preprintnumbers, letterpaper,superscriptaddress,nofootinbib]{revtex4}
\usepackage{amsfonts}
\usepackage{amsmath}
\usepackage{amssymb,epsf}
\usepackage{latexsym}
\usepackage{graphicx,epsfig}
\usepackage{amssymb}
\usepackage{subfigure}
\usepackage{epstopdf}
\usepackage[colorlinks=true,citecolor=blue,linkcolor=blue,urlcolor=black]{hyperref}

\begin{document}

\title{Novel third-order Lovelock wormhole solutions}

\author{Mohammad Reza Mehdizadeh}
\email{mehdizadeh.mr@uk.ac.ir}
\affiliation{Department of Physics, Shahid Bahonar University, P.O. Box 76175, Kerman,
    Iran}
\affiliation{Research Institute for Astronomy and Astrophysics of Maragha (RIAAM), P.O.Box 55134-441, Maragha, Iran}
\author{Francisco S. N. Lobo}
\email{fslobo@fc.ul.pt}
\affiliation{Instituto de Astrof\'isica e Ci\^encias do Espa\c{c}o, Departamento de F\'isica da Faculdade de Ci\^encias da Universidade de Lisboa, Edif\'icio C8, Campo Grande, P-1749-016 Lisbon, Portugal}

\pacs{04.50.Kd, 98.80.-k, 95.36.+x}

\date{\today}

\begin{abstract}
In this work, we consider wormhole geometries in third-order Lovelock gravity and investigate the possibility that these solutions satisfy the energy conditions. In this framework, by applying a specific equation of state, we obtain exact wormhole solutions, and by imposing suitable values for the parameters of the theory, we find that these geometries satisfy the weak energy condition in the vicinity of the throat, due to the presence of higher order curvature terms. Finally, we trace out a numerical analysis, by assuming a specific redshift function, and find asymptotically flat solutions that satisfy the weak energy condition throughout the spacetime.
\end{abstract}

\pacs{04.70.Bw, 04.30.-w, 04.70.Dy} \maketitle

\section{Introduction}

Wormholes are hypothetical geometrical shortcuts, which connect two different regions in spactime \cite{mt,vis}. A fundamental ingredient of traversable wormholes is the flaring-out condition \cite{mt}, which entails the presence of exotic matter'', i.e., the violation of the null energy conditions (NEC); in fact wormhole solutions violate all of the classical energy conditions \cite{vis}. However, it was shown that evolving wormhole geometries may present ``flashes'' of weak energy condition (WEC) violation, where the matter threading the wormhole violates the energy conditions for small intervals of time \cite{kar1,kar2,Arellano:2006ex}. A popular approach in minimizing the violation of the energy energy consists on the construction of thin-shell wormholes, where the exotic matter is concentrated at the thin-shell \cite{kis}. In fact, one can alleviate and avoid altogther the violation of the energy condition in the context of modified theories of gravity and higher dimensional theories \cite{Harko:2013yb,oli,lob1,anc,dzh}.
Among higher dimensional theories of gravity, Lovelock theory is particularly interesting. In this theory, higher order curvature terms are added to the action, which lead to second order equations \cite{lov}. Note that in wormhole physics the presence of the higher order curvature terms are particularly relevant in the vicinity of the high curvature region of the wormhole throat -- especially if one is considering a small throat radius. 

Thus, the curvature near the throat is very large and therefore the investigation of the effects of higher order curvature terms in the wormhole geometry becomes important. This has motivated an extensive research in higher-dimensional wormholes. Indeed, static wormhole solutions of second and third order Lovelock gravity have been found \cite{bha,Dehghani:2009zza}. More specifically, solutions that satisfy the energy conditions, in the vicinity of the wormhole throat, have been found in third order Lovelock gravity \cite{Dehghani:2009zza}. It was also found that the third order Lovelock term with a negative coupling constant enlarges the radius of the region of normal matter \cite{Dehghani:2009zza}.
Dynamic wormhole solutions in the framework of Lovelock gravity with compact extra dimensions that are supported by normal matter were also presented \cite{meh}.
Explicit wormhole solutions respecting the energy conditions in the whole spacetime were obtained in the vacuum and dust cases with $k=-1$, where $k$ is the sectional curvature of an $(n-2)$-dimensional maximally symmetric space \cite{mae}.
However, these solutions were further extended to the positive $k=1$ sectional curvature, where for the first time specific solutions that satisfy the weak energy
condition throughout the spacetime were found \cite{mkl}.

Specific exact solutions, consisting of vacuum static wormholes, black holes and generalized Bertotti-Robinson space-times with nontrivial torsion, were also found in eight-dimensional Lovelock theory \cite{Canfora:2008ka}. It was shown that the wormhole solution found was the first example of a smooth vacuum static Lovelock wormhole which is neither Chern-Simons nor Born-Infeld. It was also shown that the presence of torsion affected the traversability  of the wormhole for scalar and spinning particles, where torsion acted as a geometrical filter, in the sense that a large torsion increases the conditions for traversability for the scalars.
Wormhole solutions in third order Lovelock gravity with a cosmological constant term, in an $n$-dimensional spacetime $\mathcal{M}^{4}\times \mathcal{K}^{n-4}$, where $\mathcal{K}^{n-4} $ is a constant curvature space, were also extensively explored \cite{mld}. More specifically, the equations of motion were decomposed to four and higher dimensional ones, and wormhole solutions were found by considering a vacuum $\mathcal{K}^{n-4} $ space. Applying the latter constraint, the second and third order Lovelock coefficients and the cosmological constant $\Lambda$ were determined in terms of specific parameters of the model, such as the size of the extra dimensions. Using the obtained Lovelock coefficients and $\Lambda$, the $4$-dimensional matter distribution threading the wormhole was found. Furthermore, exact asymptotically flat and non-flat wormhole solutions were found. Further higher-dimensional wormhole solutions have been studied in \cite{dot,kan}.

Flat charged thin-shell wormholes of third order Lovelock gravity in higher dimensions, taking into account the cut-and-paste technique were also explored \cite{Mehdizadeh:2015dta}. Using the generalized junction conditions, the energy-momentum tensor of these solutions on the shell were determined, and the issue of the energy conditions and the amount of normal matter that supports these thin-shell wormholes were explored. The analysis showed that for negative second order and positive third-order Lovelock coefficients, there are thin-shell wormhole solutions that respect the WEC. In this case, the amount of normal matter increases as the third-order Lovelock coefficient decreases. Novel solutions were also found that possess specific regions where the energy conditions are satisfied for the case of a positive second order and negative third-order Lovelock coefficients. Finally, a linear stability analysis in higher dimensions around the static solutions was carried out, and considering a specific cold equation of state, wide range of stability regions were found.

In this work, motivated in finding solutions in third-order Lovelock gravity that satisify the energy conditions, we obtain novel wormhole geometries by considering the specific equation of state used in \cite{mkl}. We investigate the effects of the third order term of Lovelock theory and a non-constant redshift function that satisfies the WEC. More specifically, we obtain exact wormhole solutions in third order Lovelock gravity by considering a constant redshift function and show despite having normal matter at the vicinity of the throat, the WEC is generically violated. However, by considering a specific redshift function, using a numerical analysis, we present explicit solutions that satisfy the WEC throughout the spacetime.

This paper is organized as follows: In Section \ref{secII}, we present a brief review of the field equations of Lovelock gravity and their applications to the the energy conditions. In Section \ref{secIII}, we introduce an equation of state to solve the equations, and find new exact wormhole geometries and numerical solutions are also obtained. Finally, we conclude in Section \ref{secconclusion}.

\section{Action and gravitational field equations}\label{secII}

The action in the framework of third-order Lovelock gravity, is given by
\begin{equation}
I=\int d^{n+1}x\sqrt{-g}\left( \mathcal{L}_{1}+\alpha _{2}^{\prime }\mathcal{L}%
_{2}+\alpha _{3}^{\prime }\mathcal{L}_{3}\right)\,,  \label{Act1}
\end{equation}
where $\alpha_{2}^{\prime}$ and $\alpha _{3}^{\prime}$ are the second (Gauss-Bonnet) and third order Lovelock coefficients;
$\mathcal{L}_{1}=R$ is the Einstein-Hilbert Lagrangian, the term $\mathcal{L}_{2}$ is the Gauss-Bonnet Lagrangian given by
\begin{equation}
\mathcal{L}_{2}=R_{\mu \nu \gamma \delta
}R^{\mu \nu \gamma \delta }-4R_{\mu \nu }R^{\mu \nu }+R^{2},
\end{equation}
and the third order Lovelock Lagrangian $\mathcal{L}_{3}$ is defined as
\begin{eqnarray}
\mathcal{L}_{3} &=&2R^{\mu \nu \sigma \kappa }R_{\sigma \kappa \rho \tau }R_{\phantom{\rho \tau }{\mu \nu }}^{\rho \tau }+8R_{\phantom{\mu \nu}{\sigma\rho}}^{\mu \nu }R_{\phantom {\sigma \kappa} {\nu \tau}}^{\sigma \kappa }R_{\phantom{\rho \tau}{ \mu \kappa}}^{\rho \tau }
    +24R^{\mu \nu \sigma \kappa }R_{\sigma \kappa \nu \rho }R_{\phantom{\rho}{\mu}}^{\rho }+3RR^{\mu \nu \sigma \kappa }R_{\sigma \kappa\mu \nu }
    \notag \\
&&+24R^{\mu \nu \sigma \kappa }R_{\sigma \mu }R_{\kappa \nu }+16R^{\mu \nu
}R_{\nu \sigma }R_{\phantom{\sigma}{\mu}}^{\sigma }
    -12RR^{\mu \nu }R_{\mu \nu }+R^{3} \,.
\end{eqnarray}%
In Lovelock theory, for each Euler density of order $k$ in an $n$-dimensional spacetime, only terms with $k < n$ exist in the equations of motion \cite{myg}. Therefore, the solutions of the third order Lovelock theory are in $n\geq 7$ dimensions.

Varying the action (\ref{Act1}) with respect to the metric, using the convention $8\pi G_n =1$, where $G_n$ is the $n$-dimensional gravitational constant, one obtains the following gravitational field equations up to third order terms
\begin{eqnarray}
G_{\mu \nu }^{(E)}+\alpha _{2}^{\prime } G_{\mu \nu }^{(GB)}+\alpha_{3}^{\prime }G_{\mu \nu }^{(TO)}=T_{\mu \nu } \,, \label{Geq}
\end{eqnarray}
where $T_{\mu \nu}$ is the energy-momentum tensor (EMT), $G_{\mu \nu}^{E}$ is the Einstein tensor and $G_{\mu \nu}^{GB}$ and $G_{\mu \nu}^{TO}$ are given by
\begin{eqnarray*}
G_{\mu \nu }^{(GB)} &=&2(-R_{\mu \sigma \kappa \tau}R_{\phantom{\kappa \tau \sigma}{\nu}}^{\kappa \tau \sigma}-2R_{\mu \rho \nu \sigma }R^{\rho \sigma}-2R_{\mu \sigma }R_{\phantom{\sigma}\nu }^{\sigma }
    +RR_{\mu \nu })-\frac{1}{2}\mathcal{L}_{2}g_{\mu \nu } \,,
    \\
G_{\mu \nu }^{(TO)}&=&-3(4R^{\tau \rho \sigma \kappa }R_{\sigma\kappa\lambda \rho }R_{\phantom{\lambda }{\nu \tau \mu}}^{\lambda }-8R_{\phantom{\tau \rho}{\lambda \sigma}}^{\tau \rho}R_{\phantom{\sigma\kappa}{\tau \mu}}^{\sigma \kappa }R_{\phantom{\lambda }{\nu \rho \kappa}}^{\lambda }
+2R_{\nu }^{\phantom{\nu}{\tau \sigma\kappa}}R_{\sigma \kappa
\lambda \rho }R_{\phantom{\lambda \rho}{\tau \mu}}^{\lambda \rho }-R^{\tau \rho \sigma \kappa }R_{\sigma \kappa \tau \rho }R_{\nu \mu }
    \\
&&+8R_{\phantom{\tau}{\nu \sigma \rho}}^{\tau }R_{\phantom{\sigma \kappa}{\tau \mu}}^{\sigma \kappa }R_{\phantom{\rho}\kappa }^{\rho }+8R_{\phantom
{\sigma}{\nu \tau \kappa}}^{\sigma }R_{\phantom {\tau \rho}{\sigma \mu}}^{\tau \rho }R_{\phantom{\kappa}{\rho}}^{\kappa }
+4R_{\nu }^{\phantom{\nu}{\tau \sigma \kappa}}R_{\sigma \kappa\mu \rho }R_{\phantom{\rho}{\tau}}^{\rho }-4R_{\nu}^{\phantom{\nu}{\tau \sigma \kappa }}R_{\sigma \kappa \tau \rho}R_{\phantom{\rho}{\mu}}^{\rho }
+4R^{\tau \rho \sigma \kappa}R_{\sigma \kappa \tau \mu }R_{\nu \rho}
    \\
&&
+2RR_{\nu }^{\phantom{\nu}{\kappa \tau \rho}}R_{\tau \rho \kappa \mu }
+8R_{\phantom{\tau}{\nu \mu \rho }}^{\tau }R_{\phantom{\rho}{\sigma}}^{\rho }R_{\phantom{\sigma}{\tau}}^{\sigma}-8R_{\phantom{\sigma}{\nu \tau\rho }}^{\sigma }R_{\phantom{\tau}{\sigma}}^{\tau }R_{\mu }^{\rho }
-8R_{\phantom{\tau }{\sigma \mu}}^{\tau \rho }R_{\phantom{\sigma}{\tau}}^{\sigma}R_{\nu \rho }-4RR_{\phantom{\tau}{\nu \mu \rho }}^{\tau }R_{\phantom{\rho}\tau }^{\rho }
    \\&&
+4R^{\tau \rho }R_{\rho \tau }R_{\nu \mu}-8R_{\phantom{\tau}{\nu}}^{\tau}R_{\tau \rho }R_{\phantom{\rho}{\mu}}^{\rho }
+4RR_{\nu \rho }R_{\phantom{\rho}{\mu }}^{\rho }-R^{2}R_{\nu \mu })-\frac{1}{2}\mathcal{L}_{3}g_{\mu \nu }\,,
\end{eqnarray*}
respectively.

In this paper, we consider the $n$-dimensional traversable wormholes metric,
given by
\begin{equation}
ds^{2}=-e^{2\phi (r)}dt^{2}+\left[ \frac{dr^{2}}{1-\frac{b(r)}{r}}%
+r^{2}d\Omega _{n-2}^{2}\right],
  \label{WHmetric}
\end{equation}%
where $d\Omega _{n-2}^{2}$ is the metric on the surface of a $(n-2)$-sphere;
$\phi (r)$ and $b(r)$ are the redshift and shape functions, respectively \cite{mt}. The redshift function $\phi (r)$ should be finite everywhere, in order to avoid the presence of an event horizon. The shape function $b(r)$ should satisfy the flaring-out condition, which is given by $rb^{\prime}-b<0$, and should also obey $b(r)-r\leq 0$. The condition $b(r_0)=r_0$, which is the minimum value of the radial coordinate, represents the throat of the wormhole.

The EMT is given by the following diagonal form
\begin{equation}
T_{\nu }^{\mu }=\mathrm{diag} \left[ -\rho \left( r\right), \; p_{r}\left( r\right), \; p_{t}\left( r\right), \; p_{t}\left( r\right) ,...\right] \,,
  \label{def:SET}
\end{equation}
where $\rho(r)$ is the energy density and $p_{r}(r)$ and $p_{t}(r)$ are the radial and transverse pressures, respectively.
Thus, the field equation (\ref{Geq}), taking into account the metric (\ref{WHmetric}), provides the following relations
\begin{eqnarray}
\rho (r) =\frac{(n-2)}{2r^{2}}\Bigg\{-\left( 1+\frac{2\alpha_{2}b}{r^{3}}+\frac{3\alpha_{3}b^{2}}{r^{6}}\right) \frac{(b-rb^{\prime })}{r}
+\frac{b}{r}\left[ (n-3)+(n-5)\frac{\alpha_{2}b}{r^{3}}+(n-7)%
\frac{\alpha_{3}b^{2}}{r^{6}}\right] \Bigg\}, \label{rho}
\end{eqnarray}
\begin{eqnarray}
p_{r}(r) =\frac{(n-2)}{2r}\Bigg\{2\left( 1-\frac{b}{r}%
\right) \left( 1+\frac{2\alpha_{2}b}{r^{3}}+\frac{3\alpha_{3}b^{2}}{r^{6}}%
\right) \phi ^{\prime }
-\frac{b}{r^{2}}\left[ (n-3)+(n-5)\frac{\alpha_{2}b}{r^{3}}%
+(n-7)\frac{\alpha_{3}b^{2}}{r^{6}}\right] \Bigg\},  \label{tau}
\end{eqnarray}
\begin{widetext}
\begin{eqnarray}
p_{t}(r) &=&\left( 1-\frac{b}{r}\right)
\left( 1+\frac{2\alpha_{2}b}{r^{3}}+\frac{3\alpha_{3}b^{2}}{r^{6}}\right) \left[ \phi ^{\prime \prime }+{\phi ^{\prime }}^{2}+\frac{(b-rb^{\prime})\phi^{\prime}}{2r(r-b)}\right]
 -\frac{2\phi ^{^{\prime }}}{r^{4}}\left( 1-\frac{b}{r}\right) \left(
b-b^{^{\prime }}r\right) \left( \alpha _{2}+3\alpha _{3}\frac{b}{r^{3}}%
\right)
    \notag \\
&&
+\left( 1-\frac{b}{r}\right) \left( \frac{\phi^{^{\prime }}}{r}+\frac{%
    b-b^{^{\prime }}r}{2r^{2}(r-b)}\right)  \left[ (n-3)+(n-5)\frac{2\alpha _{2}b}{r^{3}}+(n-7)\frac{3\alpha_{3}b^{2}}{r^{6}}\right]
    \notag \\
&&
-\frac{b}{2r^{3}}\left[ \left( n-3\right) \left( n-4\right)+\left(
n-5\right) \left( n-6\right) \frac{\alpha_{2}b}{r^{3}}
+\left( n-7\right)
\left( n-8\right) \frac{\alpha_{3}b^{2}}{r^{6}}\right] , \label{pr}
\end{eqnarray}
where the prime denotes a derivative with respect to the radial coordinate $r$. We define $\alpha _{2}\equiv (n-3)(n-4)\alpha _{2}^{\prime }$ and $\alpha_{3}\equiv (n-3)...(n-6)\alpha _{3}^{\prime }$ for notational convenience.

In the context of the local energy conditions, we examine the WEC, which asserts $T_{\mu \nu }U^{\mu }U^{\nu}\geq 0$ where $U^{\mu }$ is a timelike vector. For the diagonal EMT (\ref{def:SET}), the WEC implies $\rho \geq 0$, $\rho +p_{r}\geq 0\ $and $\rho +p_{t}\geq 0$. Note that the last two inequalities reduce to the null energy condition (NEC). Using the field equations (\ref{rho})--(\ref{pr}), one finds the following relations
\begin{eqnarray}
\rho +p_{r}&=&-\frac{(n-2)}{2r^{2}}\left[ \frac{(b-rb^{\prime})}{r}+2\phi
^{\prime }\left( b-r\right) \right]  \left( 1+\frac{2\alpha_{2} b}{r^{3}}+\frac{3\alpha_{3}b^{2}}{r^{6}}\right) ,
\label{EGBNEC}  \\
%
\rho +p_{t}&=&-\frac{\left( b-rb^{\prime }\right) }{2r^{3}}\left( 1+\frac{
    6\alpha_{2} b}{r^{3}}+\frac{15\alpha_{3}b^2}{r^6}\right)
    +\frac{b}{r^{3}}\left[ (n-3)+(n-5)\frac{2\alpha_2 b}{r^{3}}+(n-7)\frac{3\alpha_3 b^2}{r^{6}}\right]
\notag \\&&+\phi ^{\prime }\Bigg\{\frac{b-rb^{\prime }}{2r^{2}}\left( 1+\frac{6\alpha_2 b}{r^{3}}+\frac{15\alpha_3 b^2}{r^{6}}\right)
 -\frac{b}{r^{2}}\left[(n-3)+\frac{2\alpha_2 b}{r^{3}}(n-5)+\frac{3\alpha_3 b^2}{r^{6}}(n-7)\right]
    \notag \\
&&+(n-5)\frac{2\alpha_2 b}{r^{4}}+(n-9)\frac{3\alpha_3 b^2}{r^{7}}
+\frac{1}{r}\left[(n-3)
    +\frac{2\alpha_2 b^{\prime }}{r^{2}}
+\frac{6\alpha_3 b^{\prime }b}{r^{5}}\right]\Bigg\}
    \notag \\
&& +\left( 1-\frac{b}{r}\right) \left( 1+\frac{2\alpha_2 b}{r^{3}}+\frac{3\alpha_3 b^2}{r^{6}}\right)
\left({\phi ^{\prime }}^{2}+\phi ^{\prime \prime }\right) ,
\end{eqnarray}
\end{widetext}
respectively.

One can easily show that for $\alpha_2=\alpha_3 =0$ the NEC,
and consequently the WEC, are violated at the throat, due to the flaring-out condition \cite{mt}.
Note that at the throat, Eq. (\ref{EGBNEC}) reduces to
\begin{eqnarray}
\left( \rho + p_r \right)\big|_{r=r_0}=- \frac{n-2}{2r_0^2}\left(
1-b^{\prime }_0 \right)\left( 1+\frac{2\alpha_2}{r_0^2}+\frac{3\alpha_3}{r_0^4} \right) .
   \label{WECthroat}
\end{eqnarray}
Taking into account the condition $b_0^{\prime }<1$, and for $\alpha_2 > 0$ and $\alpha_3 > 0$, one verifies the general condition $\left( \rho + p_r \right)\big|_{r=r_0}<0$. Now, for other combinations of the parameters $\alpha_2 > 0$ and $\alpha_3 > 0$, such as in Gauss-Bonnet gravity ($\alpha_3=0$) with  $\alpha_2 <0$ one may have wormhole solutions satisfying the NEC. More specifically, for the third-order Lovelock gravity, one can choose an adequate range for the parameters such that the NEC is satisfied at the throat and, in general, throughout the spacetime. In the following section, we  are interested in finding and analysing specific solutions, in particular, asymptotically flat geometries, i.e., $b(r)/r\rightarrow 0$ and $\phi (r)\rightarrow 0$ as $r\rightarrow \infty $.

\section{Specific solutions}\label{secIII}

In this section, we provide several strategies for solving the field equations.
Note that we have three equations, namely, the field equations (\ref{rho})-(\ref{pr}), with the five unknown functions $\rho (r)$, $p_{r}(r)$, $p_{t}(r)$, $b(r)$ and $\phi (r)$, respectively. To find solutions one can apply restrictions on $b(r)$ and $\phi (r)$ or on the EMT components. A common practice is to use a specific equation of state (EOS) relating the EMT components, such as, specific equations of state responsible for the present accelerated expansion of the Universe \cite{lob2} and the traceless EMT equation of state \cite{APT}.

In this work, we use a particularly interesting EOS, already explored in
\cite{mkl} and considered in \cite{APT}, given by
\begin{equation}
\rho =\omega \left[ p_{r}+(n-2)p_{t}\right] .  \label{sta}
\end{equation}%
For for $\omega=1$, it reduces to a traceless EOS, $T=0$, which is usually associated with the Casimir effect. Substituting $\rho $, $p_{r}$ and $p_{t}$ in the EOS, one obtains the following differential equation
\begin{eqnarray}
b^{\prime }(r) &=& \frac{1}{\zeta} \left[2r^{2}\omega (r-b)(r^{6}+2\alpha_2 r^3 b+3\alpha_3 b^2)\left( \phi^{\prime 2}+\phi ^{\prime \prime }\right)
+r\omega \phi ^{\prime }\eta_1 -\xi \right] ,  \label{bprim}
\end{eqnarray}
where we have defined the following parameters, for notational simplicity
\begin{eqnarray}
\eta_1 =2 r^7(n-2)-2 r^6(2n-5)b+\alpha_2\left[4(n-5)r^4 b
-2r^3(2n-11)b^2\right] +\alpha_3\left[6r(n-8)b^2-3(2n-17)b^3\right]
\notag ,
\end{eqnarray}%
\begin{eqnarray*}
\xi &=&b\Big\{\alpha_3\left[1+(n-7)\omega \right](n-10)b^2
+\alpha_2\left[1+ (n-5)\omega \right]r^3 (n-7)b+r^6\left[1+(n-3)\omega \right](n-4)\Big\},
\end{eqnarray*}%
and
\begin{eqnarray*}
\zeta&=&\omega\left[\left(15b^2-12rb\right)\alpha_3+\left(6r^3b-4r^4\right)\alpha_2+r^6\right]r^2\phi ^{\prime }
+ 3\alpha_3\left[1+(n-7)\omega\right]rb^2
    \notag\\
&& +\alpha_2\left[1+(n-5)\omega \right]2r^4b
+ r^7\left[1+(n-3)\omega \right].
\end{eqnarray*}

\subsection{Zero-tidal-force solution}

The aim of this section is to find exact wormhole solutions in third-order Lovelock gravity. Since solving the differential equation ( \ref{bprim}) is, in general, too complicated, we will consider restrictions on the redshift function. Thus, in order to simplify the analysis, we will consider a zero redshift function $\phi(r)=0$  in Eq. ( \ref{bprim}), which corresponds to a vanishing tidal force \cite{mt}. 
Applying this choice, and taking into account that Eq. ( \ref{bprim}) is a rational ordinary differential equation with symmetries, one can build an integrating factor, so that the shape function can be written as 
\begin{eqnarray}
\left\{[1+(n-7)\omega]\alpha_3\right\} b^3+\left\{\alpha_2r^3[1+(n-5)\omega]\right\}b^2
+\left\{r^6[1+(n-3)\omega]\right\}b-c_2r^{10-n}=0,
\label{cubic1}
\end{eqnarray}
where the integration constant $c_2$ can be determined from the boundary condition $b(r_0)=r_0$ at the throat, and is given specifically by
\begin{eqnarray}
c_2&=&[1+(n-7)\omega]r_0^{n-7}\alpha_3+[1+(n-5)\omega]r_0^{n-5}\alpha_2
+[1+(n-3)\omega]r_0^{n-3}\,.
\end{eqnarray}

By solving the cubic equation, the general solution of Eq. (\ref{cubic1}) is given by
\begin{eqnarray}
b(r)=-\frac{[1+(n-5)\omega]\alpha_2 r^3}{3[1+(n-7)\omega]\alpha_3}+\delta+u\delta^{-1},
\end{eqnarray}
where
\begin{eqnarray}
\delta&=&\left(v+\sqrt{v^2-u^3}\right)^{1/3} ,
   \notag  \\
v&=&\frac{c_2 r^{10-n}}{2\alpha_3\eta}+\frac{[1+(n-5)\omega][1+(n-3)\omega]\alpha_2r^9}{6 \alpha_3^2 \eta^2}
-\frac{[1+(n-5)\omega]^{3}\alpha_2^3r^{27}}{27\alpha_3^3\eta^3} ,
     \notag \\
u&=&\frac{[1+(n-5)\omega]^{2}\alpha_2^2r^6}{9{\alpha_3}^2\eta^2}
-\frac{3[1+(n-3)\omega][1+(n-7)\omega]r^6}{9\alpha_3 \eta^2} ,\notag
\end{eqnarray}
and $\eta=[1+(n-7)\omega]$ are defined for notational simplicity.

The flaring-out condition, at the throat, obeys the following inequality
\begin{eqnarray}
b^{\prime}(r_0)=-\frac{[1+(n-3)\omega](n-4)r_0^4+\xi_0}{[1+(n-3)\omega]r_0^4+\xi_1}
< 1 \,,
\end{eqnarray}
where
\begin{eqnarray}
\xi_0&=&\alpha_2[1+(n-5)\omega](n-7)r_0^2
+\alpha_3 [1+(n-7)\omega](n-10)\,,
   \notag \\
\xi_1&=&2\alpha_2[1+(n-5)\omega]r_0^2+
3\alpha_3[1+(n-7)\omega]\,.  \notag 
\end{eqnarray}

The EMT profile for this solution are given by
\begin{align}
\rho=&-\frac{(n-2)(n-1)\Sigma_1\omega b^2}{r^6\Gamma},
	\\
\rho+p_r=&-\frac{(n-2)\left(3\alpha_3 b^2+2\alpha_2 b r^3+r^6\right)b\,\Sigma_2}{2r^9\Gamma} ,
	\\
\rho+p_t=&\frac{b\,\Sigma_3}{2r^9\Gamma},
\end{align}
where
\begin{eqnarray}
\Sigma_1&=&\left(4r^3\alpha_3b+r^6\alpha_2+\alpha_2\alpha_3b^2 \right),
    \notag\\
\Sigma_2 &=&\left[(n-7)\alpha_3[1+\omega(n-7)]\right]b^2
  +r^3\alpha_2(n-5)[1+\omega(n-5)]b
  +r^6(n-3)[1+\omega(n-3)] ,
    \notag\\
\Sigma_3 &=&3(n-7)\alpha_3^2[1+(n-7)\omega]b^4
+3[(n^2-18n+57)\omega-9+n]\alpha_3r^3\alpha_2b^3
    \notag\\&&
+2\{[(\alpha_2^2-2\alpha_3)n^2+(-8\alpha_3-10\alpha_2^2)n
+25\alpha_2^2+34\alpha_3]\omega+
(\alpha_2^2-2\alpha_3)n-4\alpha_3-5\alpha_2^2 \}r^6b^2
    \notag\\&&
+[(-18n+41+n^2)\omega-9+n]r^9
\alpha_2 b
+[1+(n-3)\omega]r^{12}(n-3) , \notag
\end{eqnarray}
and
\begin{eqnarray}
\Gamma &=&3\alpha_3(1+\omega(n-7))b^2
+  2\alpha_2r^3[1+\omega(n-5)]b
+r^6[1+\omega(n-3)]\,. \notag
\end{eqnarray}
Note that for the cases $\alpha_2>0$ and $\alpha_3>0$, we have the general condition $\rho+p_r<0$ at the throat (as mentioned above), which is readily verified as the factor $\left( 1+2\alpha_{2} b/r^{3}+3\alpha _{3}b^{2}/r^{6}\right)$ in Eq. (\ref{EGBNEC}), is positive.
Thus, in order to satisfy the WEC, we should search for solutions where either one of the parameters $\alpha_2$ and $\alpha_3$, or both, are negative. Indeed, this is possible in a specific range of the radial coordinate, in the vicinity of the wormhole throat.
For instance, one can choose suitable values of $\alpha_2$ and $\alpha_3$ such that $\rho$ and $\rho+p_t$ have no real root and therefore are positive everywhere, while $\rho+p_r$ possesses a real root ($r_c$), where the value $\rho+p_r$ is positive in the radial region $r_{0}\leq {r}\leq {r_{c}}$, signalling normal matter, where  $r_c$  corresponds to the positive real roots of the equation
\begin{equation}
r^6+2\alpha_2 r^3 b +3\alpha_3 b^2=0,  \label{rmax}
\end{equation}
which follows from Eq. (\ref{EGBNEC}).
In addition to this,  Eq. (\ref{WECthroat}) entails a choice of the Lovelock coefficients where $r_{-}<r_0<r_{+}$, where
\begin{equation}
r_{\pm}=\left( -\alpha _{2} \pm \sqrt{\alpha _{2}^{2}-3\alpha _{3}}\right) ^{1/2}
.  \label{r+}
\end{equation}

\begin{figure}[tbp]
    \begin{center}
    \subfigure[]
    {\label{wor2a}
    \includegraphics[width=0.48\textwidth,height=0.3\textheight]{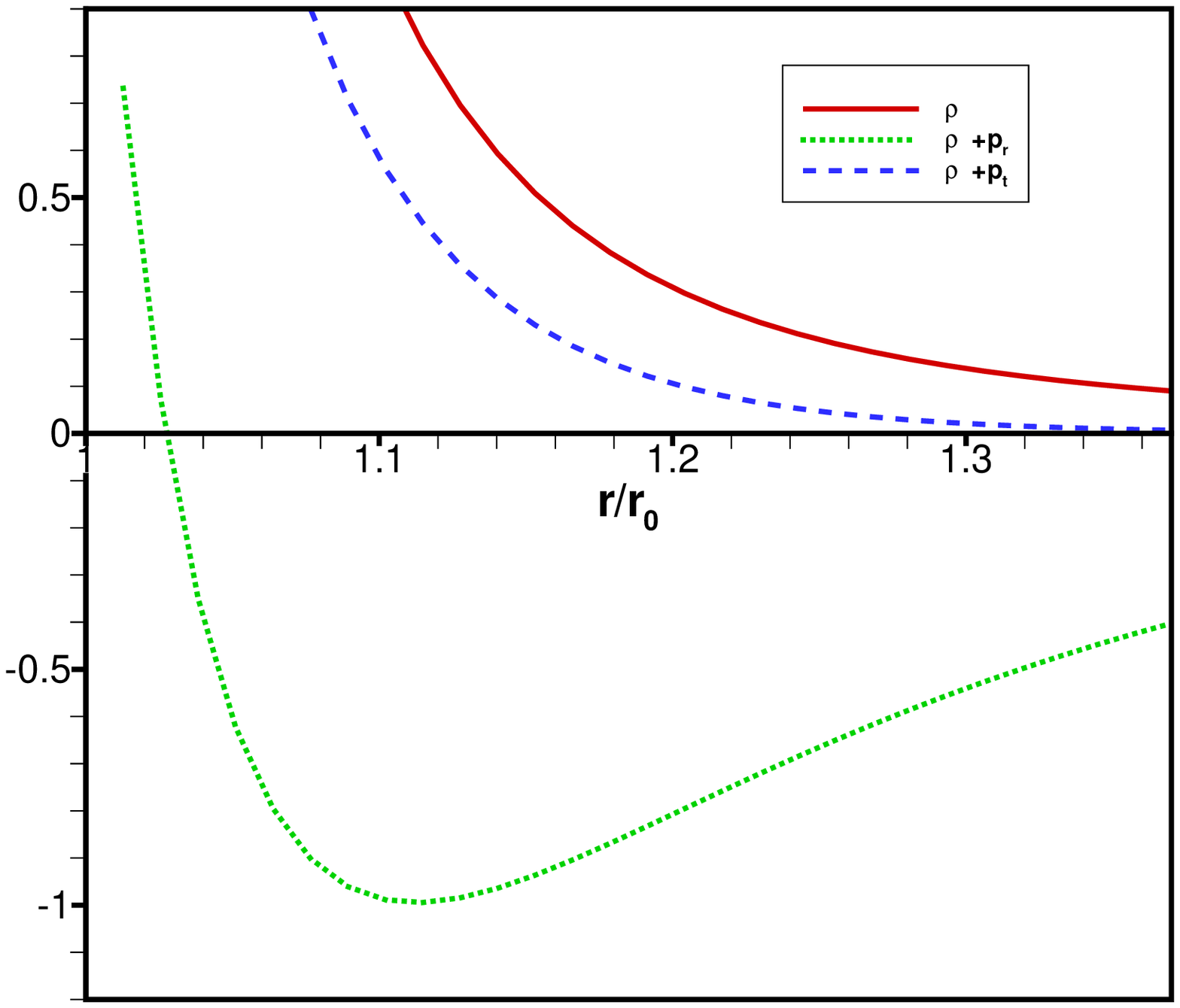}}
    \text{\hspace{0cm}}
    \subfigure[]
    {\label{wor2b}
    \includegraphics[width=0.48\textwidth,height=0.3\textheight]{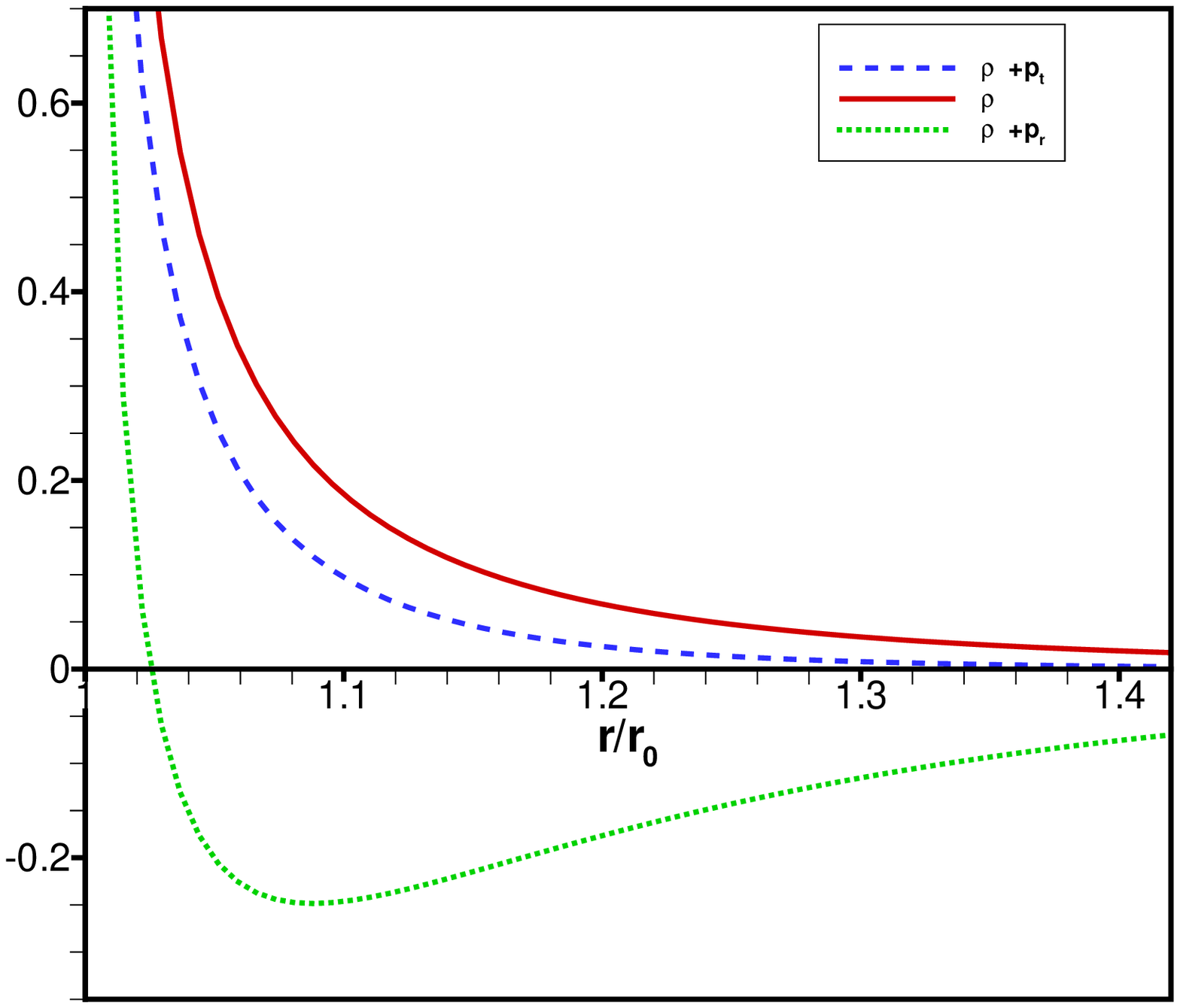}}
    \end{center}
    \caption{The behavior of $\protect\rho $ (solid), $\protect\rho +p_{r}$ (dotted) and $\protect\rho +p_{t}$ (dashed) versus $r/r_{0}$ for  $\protect\omega =1$ ($T=0$), $n=7$. The constants are $\alpha_2=-7.2,0.6$, $\alpha_3=4,-4$ and $r_0=3,1.92$,  respectively, in the left (a) and right (b) plots, respectively.}
    \label{wor2}
\end{figure}

We plot the quantities $1-b(r)/r$, $\rho$, $\rho+p_r$ and $\rho+p_t$ in Fig. \ref{wor2}. The components of the EMT tend to zero as $r$ tends to infinity. We have considered $n=7$ in both plots, for the specific case of $\omega=1$, which reduces to a traceless EMT, with $T=0$. For the Fig. \ref{wor2a}, we have considered $\alpha_2=-7.2$, $\alpha_3=4$, and in plot \ref{wor2b}, $\alpha_2=0.6$, $\alpha_3=-4$, respectively. The plots show that it is possible to choose suitable values for the constants in order to have normal matter in the vicinity of the throat. One can also see from Eq. (\ref{rmax}) that the radius of the region of normal matter increases as $\alpha_3$ becomes more negative.

\subsection{Numerical solutions satisfying the WEC}

In this section, we find asymptotically flat solutions, where the WEC is satisfied throughout the spacetime. As mentioned above, since analytic solutions are extremely difficulty to find, we adopt a numerical approach in solving Eq. (\ref{bprim}). For this purpose, we choose an asymptotically flat redshift function given by
\begin{equation}
\phi (r)= \frac{\phi_1}{2}\left( \frac{r_{0}}{r}\right) ^{m},
\label{phii}
\end{equation}
where $\phi_{1}$ is a dimensionless constant and $m$ is a positive constant. This choice guarantees that the redshift function is finite everywhere.

Now, solving Eq. (\ref{bprim}) numerically, we choose constant parameters so that solutions are asymptotically flat. Recall that for the cases that either of $\alpha_2$ and $\alpha_3$, or both, are negative, one can in principle construct wormhole solutions that satisfy the WEC at the wormhole throat, so that with choices for these parameters one may obtain normal matter in limit of large $r$, and thus satisfy the WEC throughout the spacetime.

\begin{figure}[tbp]
    \begin{center}
    \subfigure[]
    {\label{k0a}
    \includegraphics[width=0.48\textwidth,height=0.28\textheight]{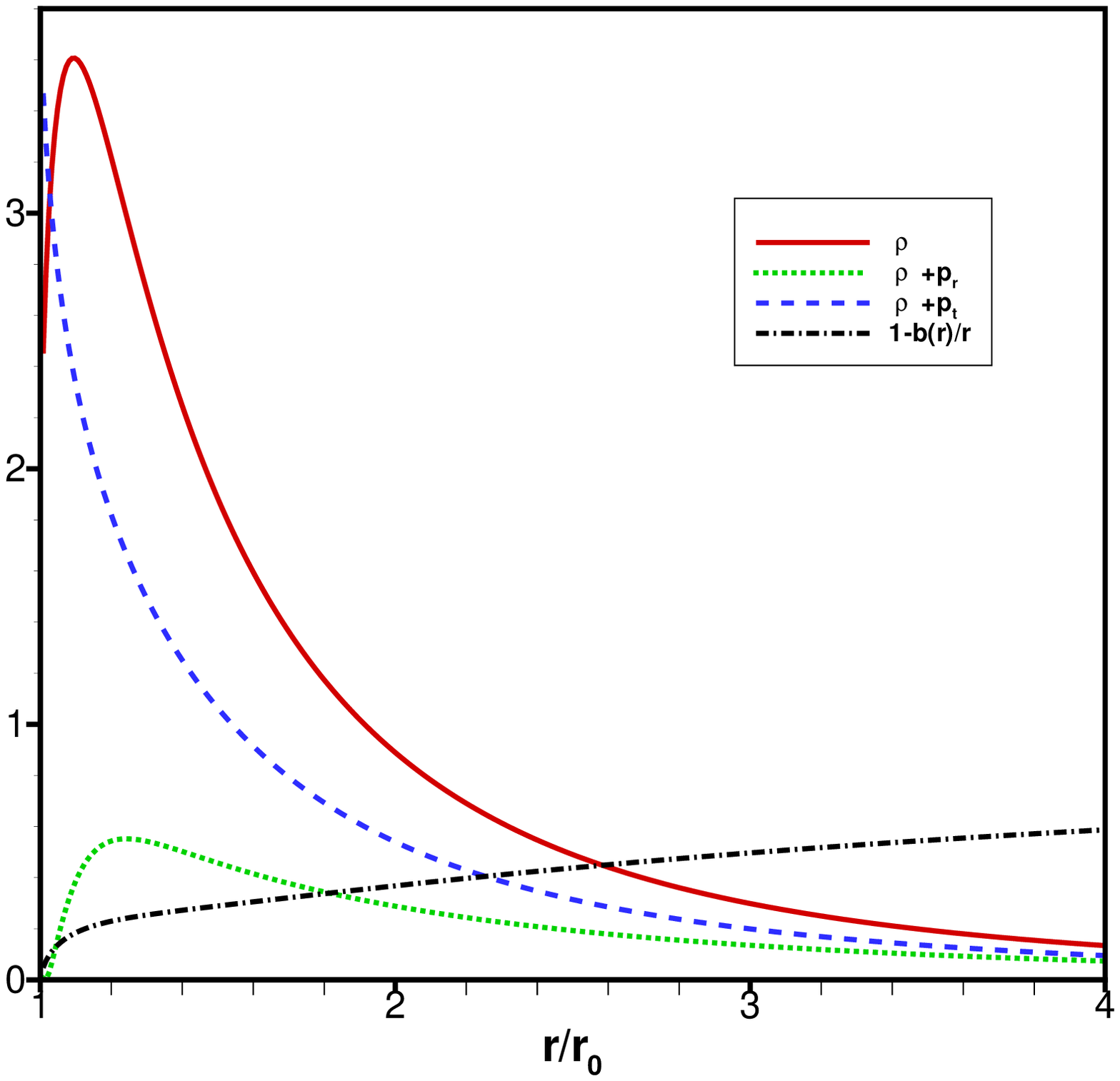}}
    \text{\hspace{0cm}}
    \subfigure[]
    {\label{k1a}
    \includegraphics[width=0.48\textwidth,height=0.28\textheight]{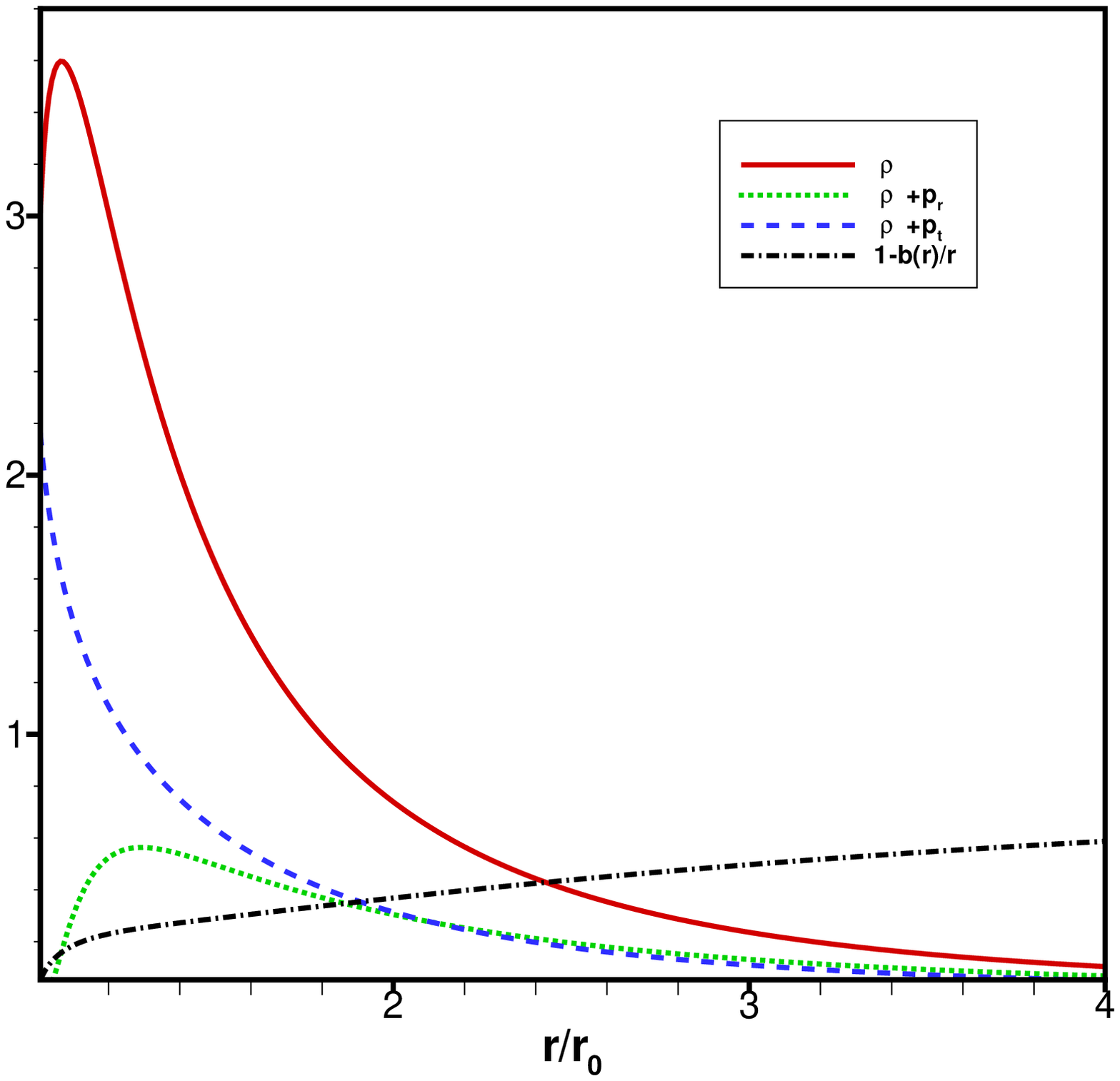}}
    \text{\hspace{0cm}}
    \subfigure[]
    {\label{k3a}
    \includegraphics[width=0.48\textwidth,height=0.28\textheight]{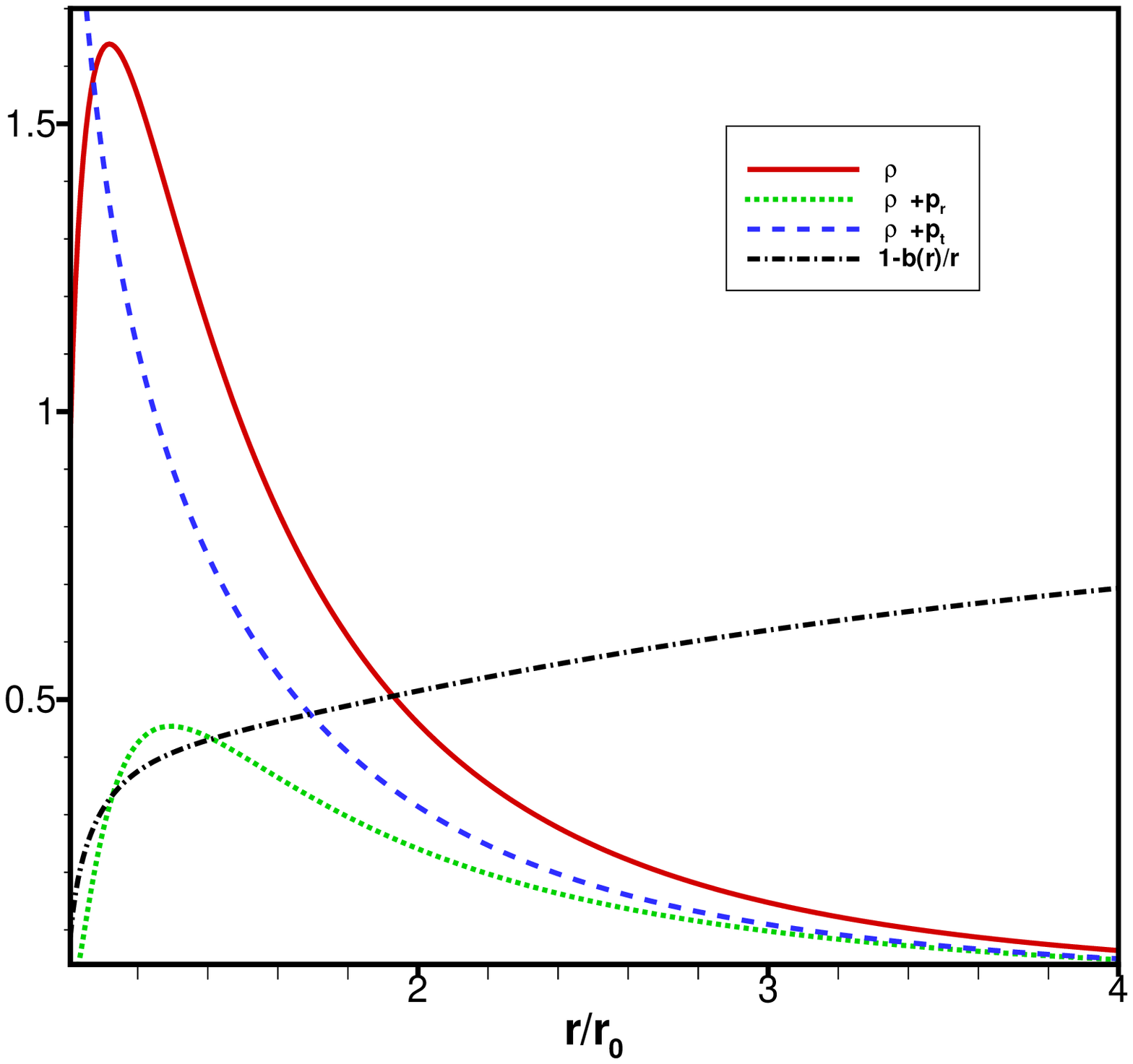}}
    \end{center}
 \caption{The behavior of $1-b(r)/r$ (dotted-dashed), $\protect\rho $ (solid),
    $\protect\rho +p_{r}$ (dotted) and $\protect\rho +p_{t}$ (dashed) versus $r/r_{0}
    $ for $\protect\omega =1$ ($T=0$), $n=7$. The constants are $\alpha_2=-.
    45,0.2,-2$, $\alpha_3=-1.6,-3.2,0.7$, $\phi_1=-10.9,-8,-6$ and $r_0=1.58, 1.58,
    1.7$, respectively, for the top left (a), top right right (b) and bottom (c) 
    plots, respectively. This solution satisfies the WEC throughout the entire 	
    spacetime.}
    \label{ka}
 \end{figure}

As in the previous section, we consider $n=7$, and for $\omega=1$ and, so that $T=0$. We have chosen the following values for the parameters: In Fig. \ref{k0a}, $\alpha_2=-0.45$, $\alpha_3=-1.6$, $\phi_1=-10.9$; in Fig. \ref{k1a}, $\alpha_2=0.2$, $\alpha_3=-3.2$, $\phi_1=-8$; and in Fig. \ref{k3a}, $\alpha_2=-2$, $\alpha_3=0.7$, $\phi_1=-6$. We verify that for these choices, the quantity $b(r)/r$ tends to zero at spatial infinity. For these choices, the quantities $\rho(r)$, $\rho(r)+p_r(r)$ and $\rho(r)+p_t(r)$ are positive throughout the spacetime, implying that the WEC is satisfied $\forall r$. These results are explicitly depicted in Fig. \ref{ka}.

\section{Summary and Conclusion}\label{secconclusion}

In this paper, we have explored higher-dimensional wormhole solutions of third order Lovelock gravity by considering specific choices for the redshift function and by imposing a particular equation of state. More specifically, we obtained exact wormhole solutions in third order Lovelock gravity by considering a constant redshift function and showed that for the cases that either of $\alpha_2$ and $\alpha_3$, or both of them, are negative, one can obtain a region with normal matter near the throat. It was also shown that the radius of the region with normal matter near the wormhole throat enlarges as $\alpha_3$ becomes more negative. In the context of Gauss Bonnet gravity, we found solutions that satisfied the WEC throughout the entire spacetime \cite{mkl}. 
These solutions were obtained by considering a negative Gauss-Bonnet coupling constant, i.e., $ \alpha_2 < 0$. We have extended this analysis to third-order Lovelock gravity, in this paper, by finding solutions that satisfy the WEC throughout the entire spacetime where either of $\alpha_2$ and $\alpha_3$, or both, are negative.

\acknowledgments{MRM has been supported financially by the Research Institute for Astronomy \& Astrophysics of Maragha (RIAAM)under research project No.1/3252-7, Iran. FSNL acknowledges financial support of the Funda\c{c}\~{a}o para a Ci\^{e}ncia e Tecnologia through an Investigador FCT Research contract, with reference IF/00859/2012, and the grants PEst-OE/FIS/UI2751/2014 and UID/FIS/04434/2013. }



\end{document}